\documentclass{appolb}
\usepackage{epsfig}

\newcommand {\pom} {I\!\!P}
\newcommand {\pomsub} {{\scriptscriptstyle \pom}}
\newcommand {\reg} {I\!\!R}
\newcommand {\regsub} {{\scriptscriptstyle \reg}}

\newcommand {\apom} {\alpha_{\pomsub}}
\newcommand {\areg} {\alpha_{\regsub}}

\begin{document}

\title{Cross sections at HERA%
\thanks{Talk presented at the 2009 Epiphany meeting, Cracow, Jan 5-7, 2009.}
}
\author{Aharon Levy
\address{The Raymond and Beverly Sackler School of Physics and Astronomy,
Tel Aviv University, 69978 Tel Aviv, Israel}
}
\maketitle

\begin{abstract}
The energy dependence of cross sections is discussed as a way to learn
about the dynamics of photon induced interactions at HERA. The
question of determining the scale of exclusive electroproduction of
vector mesons is addressed.
\end{abstract}

\vspace{0.15cm}
This talk is dedicated to the memory of Jan Kwiecinski.

\section{Introduction}

The energy dependence of a given process teaches us about the dynamics
of this process. If the energy dependence can be described by a
Regge-type approach a la Donnachie and Landshoff~\cite{dl} we call the
process soft. The total hadron-proton cross sections dependence on
center-of-mass energy exemplifies the behaviour of soft processes.. If
the energy dependence can be described by perturbative QCD (pQCD) we
say that the process is hard. An example is exclusive $J/\psi$
electroproduction~\cite{zjpsi,hjpsi}. The nice feature of $ep$
interactions at HERA is that we can see the interplay between soft and
hard interactions~\cite{afs}.

Jan Kwiecinski made very many and important contributions to this
subject by studying actually all possible reactions, like
hadron-hadron ($hh$), $\gamma p$, $\gamma \gamma$ and $e p$. Many of the
studies that we have been doing at HERA were inspired by his
findings. In this sense it is very appropriate that this conference is
dedicated to his memory.

\begin{figure}[htb]
\includegraphics[width=0.5\hsize]{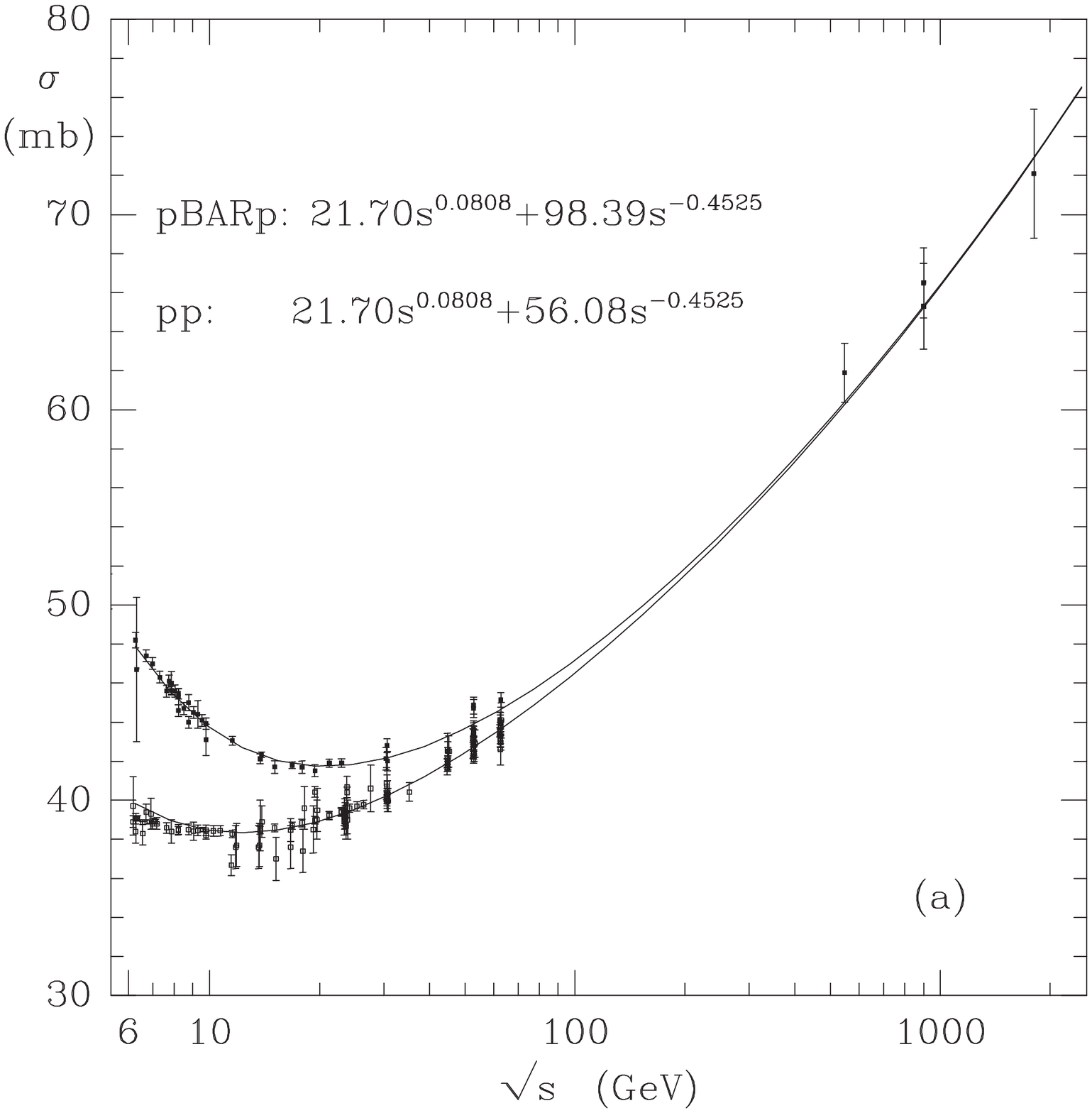}
\includegraphics[width=0.5\hsize]{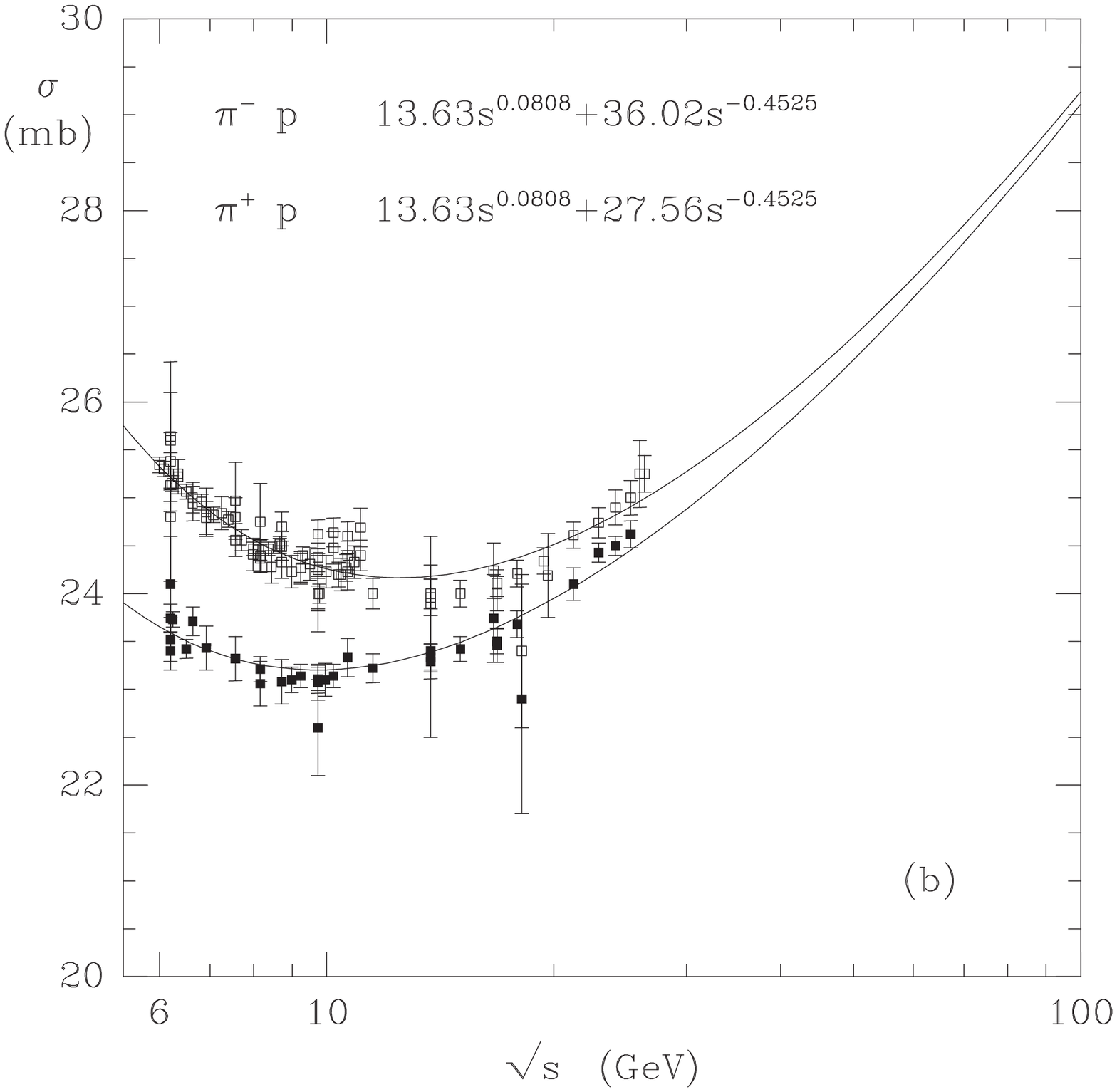}
\vspace*{-2.cm}
\caption{
The total cross section data of $p p$, $\bar{p}p$, $\pi^+p$ and
$\pi^-p$ interactions. The curves are the DL fits to~(\protect\ref{eq:dl}).  }
\label{fig:dl}
\end{figure}

What sets the baseline for soft processes? Donnachie and Landshoff
(DL) showed~\cite{dl} that all $hh$ total cross sections can be
described by a simple expression containing two terms, inspired by
Regge theory~\cite{regge}, one term representing the contribution of
the effective Pomeron trajectory and a second term of a Regge
trajectory,
\begin{eqnarray}
\sigma_{tot}(hh) &=& A s^{\epsilon} + B s^{\eta}\\
                  &=& A s^{0.0808} + B s^{-0.4525}, 
\label{eq:dl}
\end{eqnarray}
where $s$ is the square of the center-of-mass energy, $A$ and $B$ are
process-dependent constants, $\epsilon = \apom(0)-1, \eta =
\areg(0)-1$, and $\apom(0) (\areg(0))$ is the intercept of the
effective Pomeron (Reggeon) trajectory. An example of $s$ dependence
of the total cross sections for $\bar{p}p, pp, \pi^+p$ and $\pi^-p$
interactions is shown in Fig.~\ref{fig:dl}, together with the DL fits.

Thus, at high energies, the expected energy dependence for soft
interactions is $s^{0.08}$ or $W^{0.16}$, where $W$ is the
center-of-mass energy. A more recent fit to the total cross section
data~\cite{cudell} results in a value $\epsilon = 0.096$.

\begin{figure}[htb]
\begin{center}
\includegraphics[width=0.5\hsize]{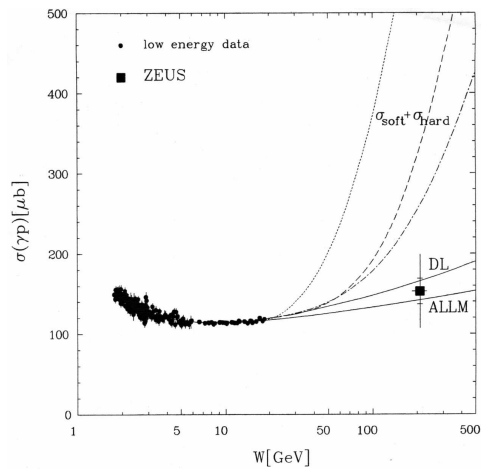}
\caption{
The first total $\gamma p$ cross section measured by ZEUS at
HERA~\protect\cite{sigtot92}, together with low energy data. The curves are
different predictions for the $W$ behaviour of the cross section,
before HERA started running.  } \label{fig:sigtot92}
\end{center}
\end{figure}

When HERA started running there were quite extreme predictions for the
value of the total photoproduction cross sections. They ranged from
150 $\mu$b, compatible with the predictions of Donnachie and Landshoff,
expected from a soft process, to values as high as 1 mb, coming from
hard components which dominate the process. Seven seconds of running
and a handful of events were enough to establish~\cite{sigtot92} that
the photon behaves like a hadron at HERA energies, as seen in
Fig.~\ref{fig:sigtot92}.
The mild increase with energy indicated that
the dominant configuration of the photon is its fluctuation in a
large-size $q\bar{q}$ pair.

\section{Exclusive vector meson photoproduction}

\begin{figure}[htb]
\begin{center}
\includegraphics[width=0.6\hsize]{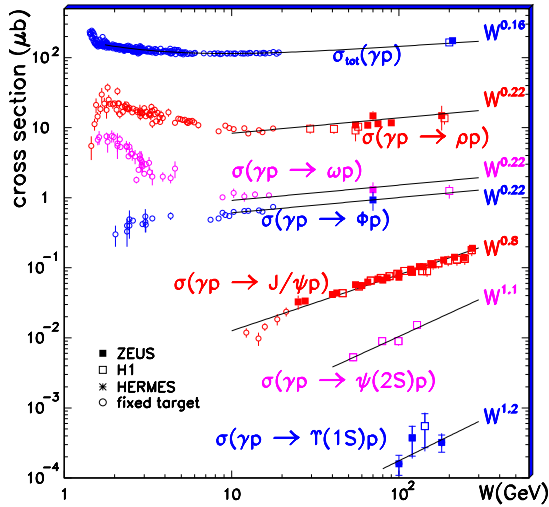}
\vspace*{-0.1cm}
\caption{
Total and exclusive vector meson photoproduction data, as a function
of $W$. The curves are fits of the form $\sim W^\delta$.  }
\label{fig:sigvm}
\end{center}
\end{figure}

The total photoproduction cross section is plotted in
Fig.~\ref{fig:sigvm}, together with the cross sections for exclusive
photoproduction of vector mesons, $\sigma(\gamma p \to V p)$, as a
function of $W$. 
One can describe the high energy part of the cross section by the form
$W^\delta$, where for the total cross section $\delta = 2
\epsilon$. The value of $\delta$ for the light vector mesons $\rho,
\omega$ and $\phi$ is also consistent with a soft process; here too the
photon fluctuates into a pair of light $q\bar{q}$ quarks and thus is
in a large configuration. However, in case of the $J/\psi$, the pair
of heavy quarks squeeze the photon into a small configuration, there
is color screening and therefore the cross section is low. However,
the pair being small, it can resolve the partonic structure of the
proton. As the reaction is exclusive, the cross section is
proportional to the gluon density squared. Since the gluon density
rises strongly with decreasing Bjorken-$x$, meaning increasing $W$,
one gets a strong energy dependence in the case of the $J/\psi$. As
the scale gets larger, where in this case the scale is set by the mass
of the vector meson, the gluon density is steeper, resulting in a
steeper $W$ dependence. Thus the behaviour seen in
Fig.~\ref{fig:sigvm} is a nice manifestation of the transition from
soft to hard processes.

\section{Exclusive vector meson electroproduction}

In the earlier section, the scale was set by the vector meson mass. We
can now fix the mass of the vector meson and use the virtuality $Q^2$
of the photon as the scale. In this section we discuss the exclusive
electroproduction of $\rho^0$~\cite{zeusrho,hrho},
$\phi$~\cite{zphi,hphi} and $J/\psi$~\cite{zjpsi,hjpsi}. In addition
we also look at deeply virtual Compton scattering
(DVCS)~\cite{zdvcs,hdvcs}, namely $\gamma^* p \to \gamma p$.

\begin{figure}[htb]
\hspace{-1.0cm}
\includegraphics[width=0.55\hsize]{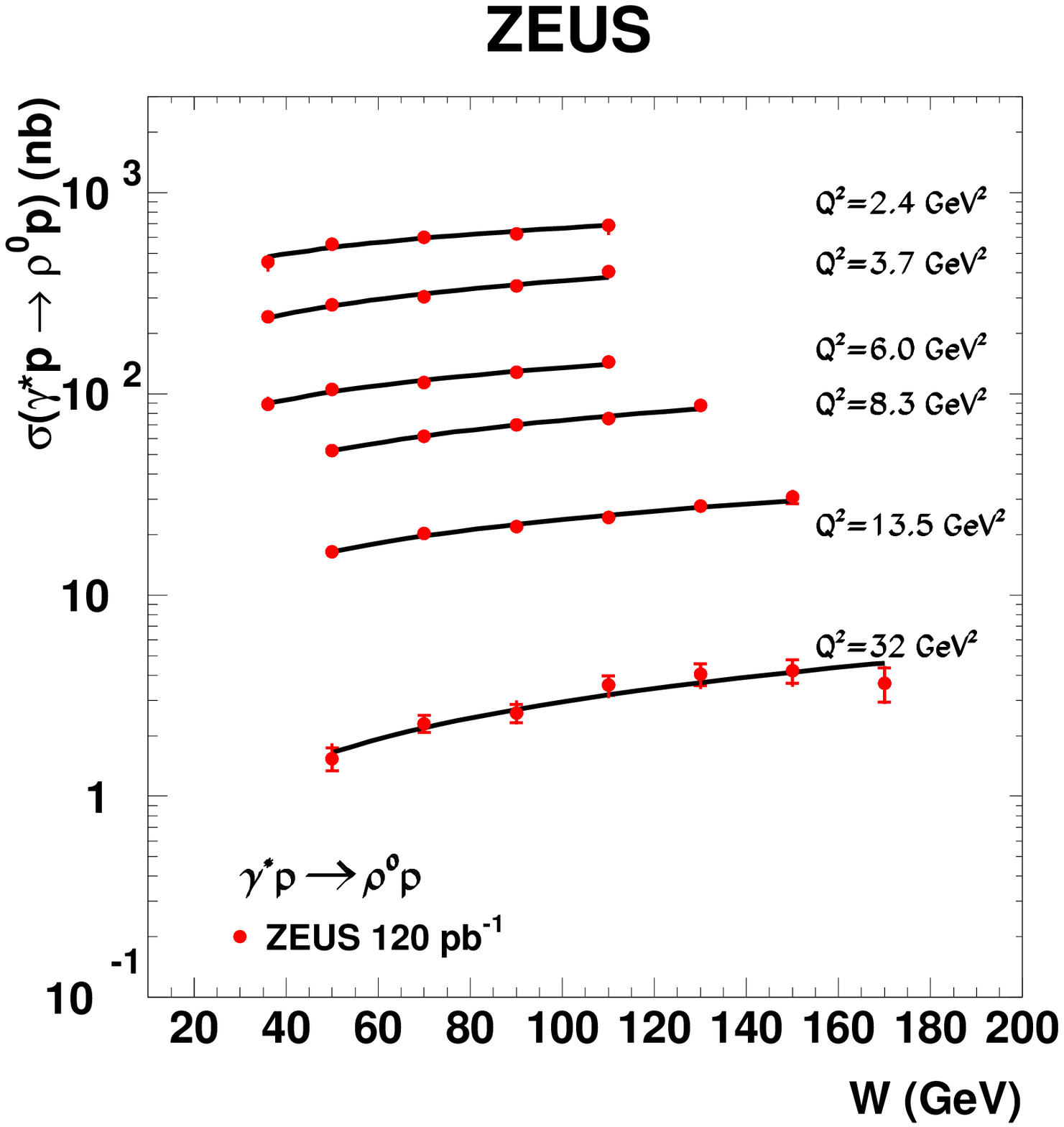}
\includegraphics[width=0.55\hsize]{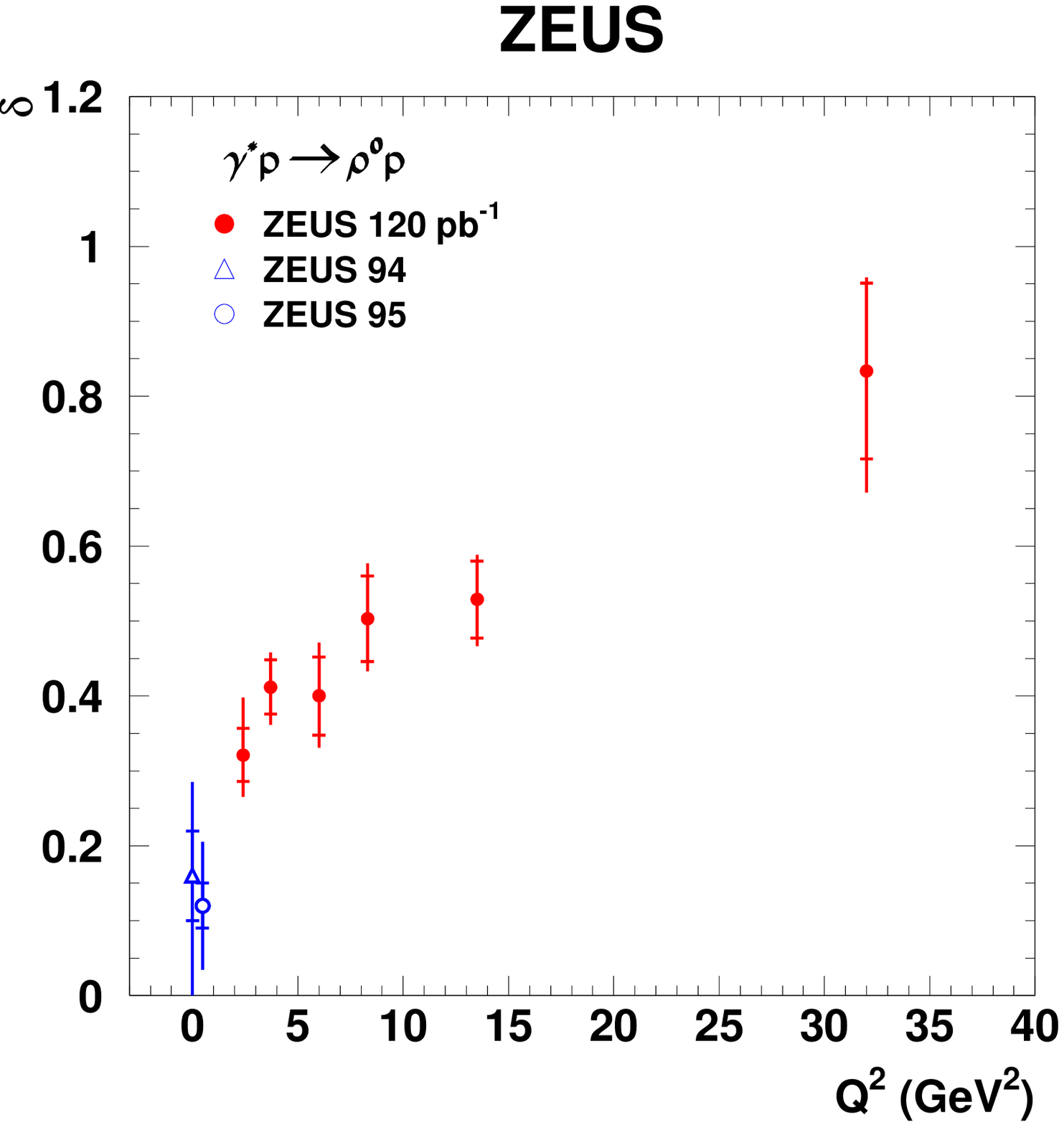}
\caption{
(Left-hand side) The $W$ dependence of the cross section for $\gamma^* p
\to \rho^0 p$ at different $Q^2$ values. The curves are fits of the
form $\sim W^\delta$.(Right-hand side) The $Q^2$ dependence of $\delta$. }
\label{fig:sig-rho}
\end{figure}

The cross section $\sigma(\gamma^*p\to\rho^0 p$)~\cite{zeusrho} is
presented in Fig.~\ref{fig:sig-rho} as a function of $W$, for
different values of $Q^2$. The cross section rises with $W$ at all
$Q^2$ values. In order to quantify this rise, the logarithmic derivative
$\delta$ of $\sigma$ with respect to $W$ is obtained by fitting the
data to the expression $\sigma\propto W^\delta$ for each $Q^2$
value. The resulting values of $\delta$ are shown in
Fig.~\ref{fig:sig-rho} and display a clear and significant rise from a
value of $\sim$ 0.1-0.2 at low $Q^2$, as expected for soft processes,
to that of $\sim$ 0.8 at higher $Q^2$, which is consistent with twice
the logarithmic derivative of the gluon density with respect to $W$.

\begin{figure}[htb]
\hspace{-0.5cm}
\includegraphics[width=0.35\hsize]{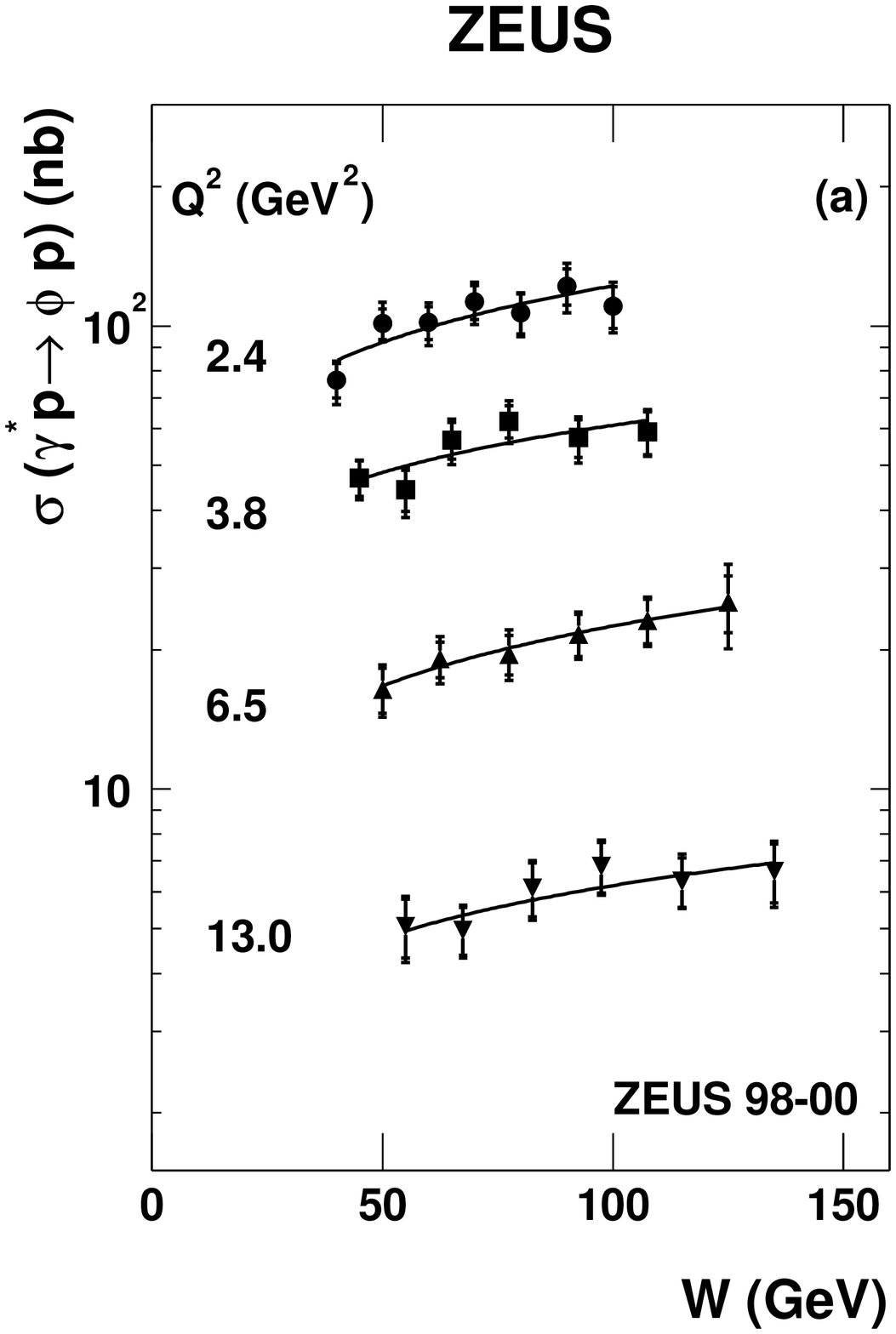}
\includegraphics[width=0.55\hsize]{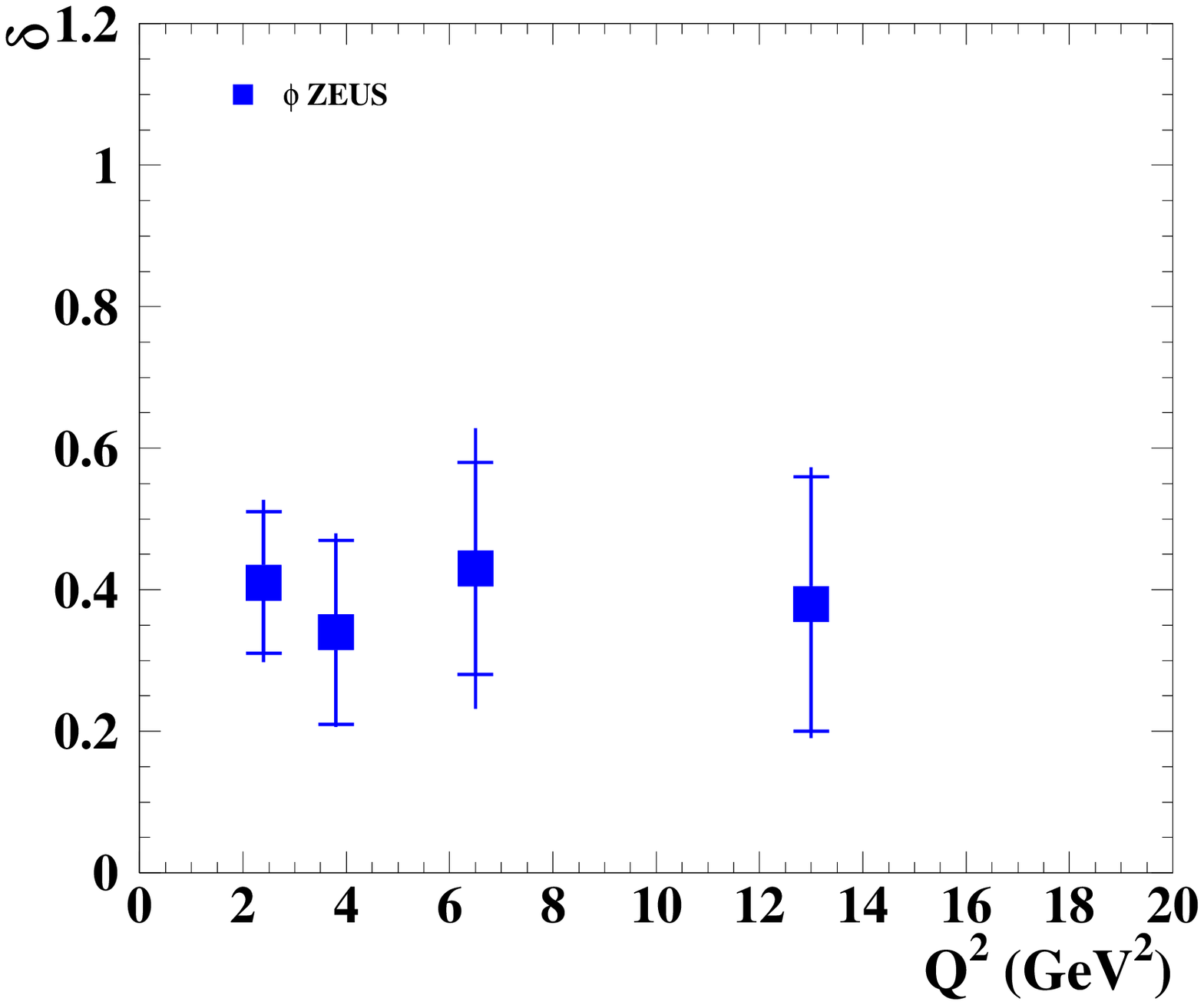}
\vspace*{-1.0cm}
\caption{
(Left-hand side) The $W$ dependence of the cross section for $\gamma^* p
\to \phi p$ at different $Q^2$ values. The curves are fits of the
form $\sim W^\delta$.(Right-hand side) The $Q^2$ dependence of $\delta$. }
\label{fig:sig-phi}
\end{figure}

An increase of the cross section $\sigma(\gamma^*p\to\phi p$) with $W$
for different $Q^2$ values is also observed, as displayed in
Fig.~\ref{fig:sig-phi}. However, the data are not precise enough to
see a change of the slope with $Q^2$.

\begin{figure}[htb]
\hspace{-1.0cm}
\includegraphics[width=0.40\hsize]{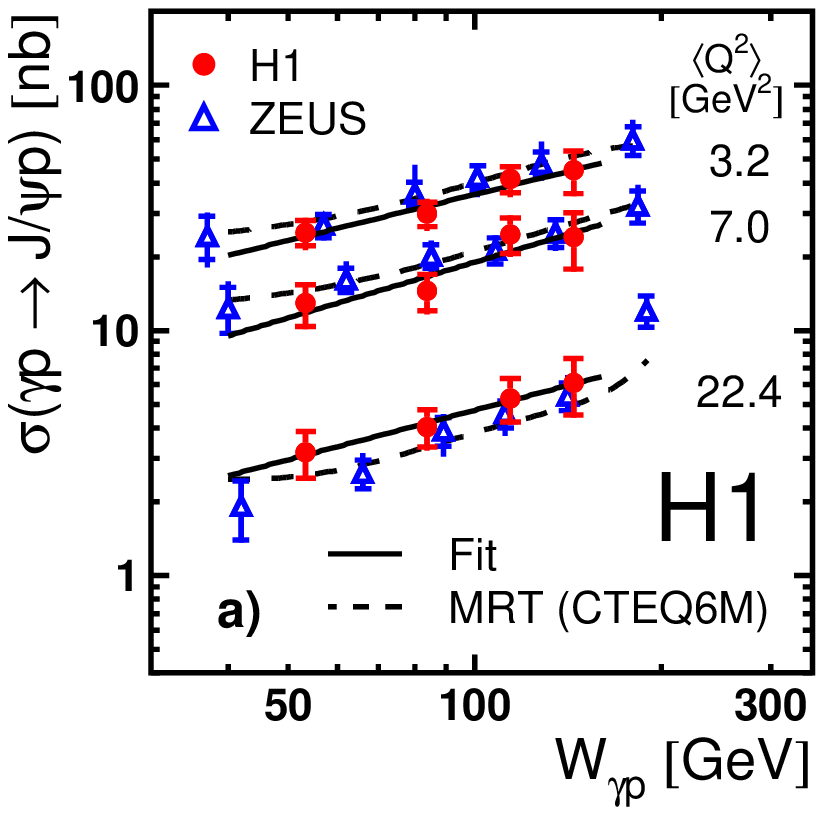}
\includegraphics[width=0.65\hsize]{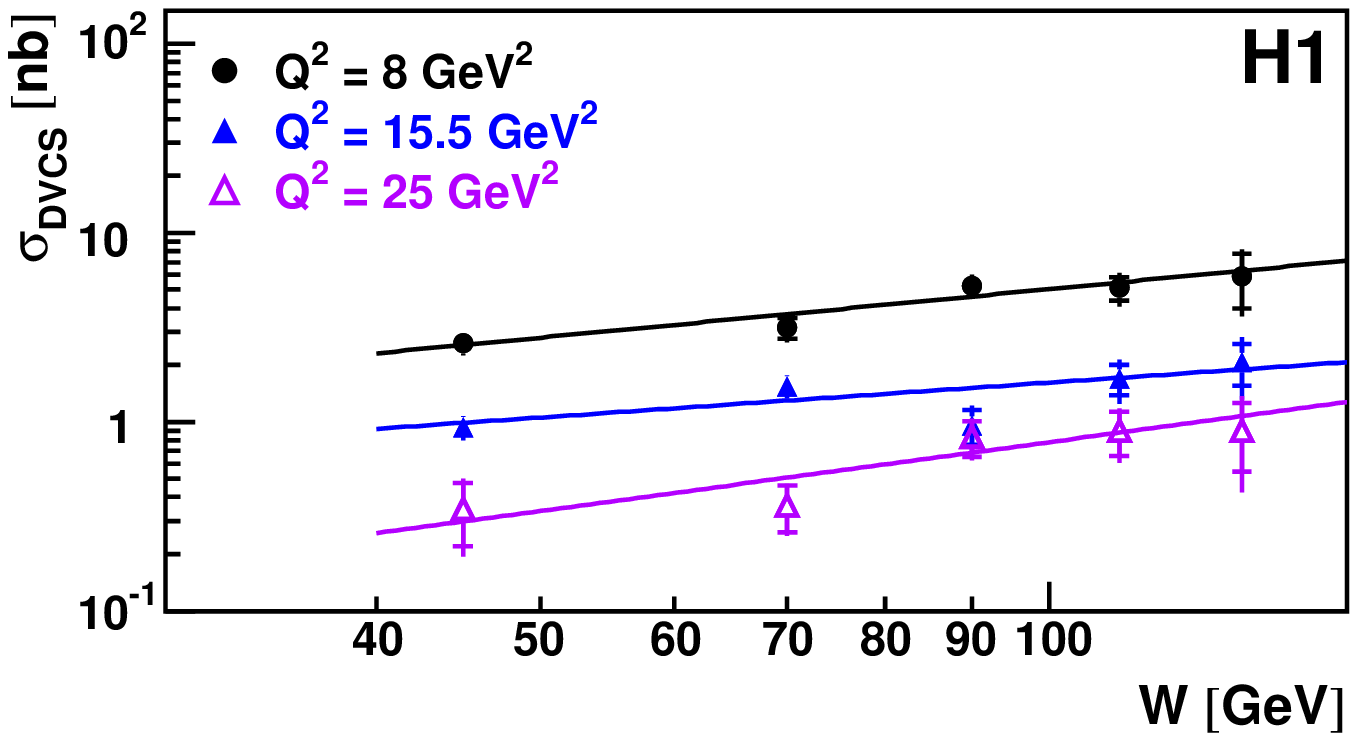}
\vspace*{-1.cm}
\caption{
(Left-hand side) The $W$ dependence of the cross section for $\gamma^* p
\to J/\psi p$ at different $Q^2$ values.
The curves are fits of the form $\sim W^\delta$.  (Right-hand side)
The $W$ dependence of the cross section for $\gamma^* p
\to \gamma p$ at different $Q^2$ values, measured by the H1 collaboration.
The curves are fits of the form $\sim W^\delta$.}
\label{fig:sig-dvcs}
\end{figure}

\begin{figure}[htb]
\hspace{-1.0cm}
\includegraphics[width=0.55\hsize]{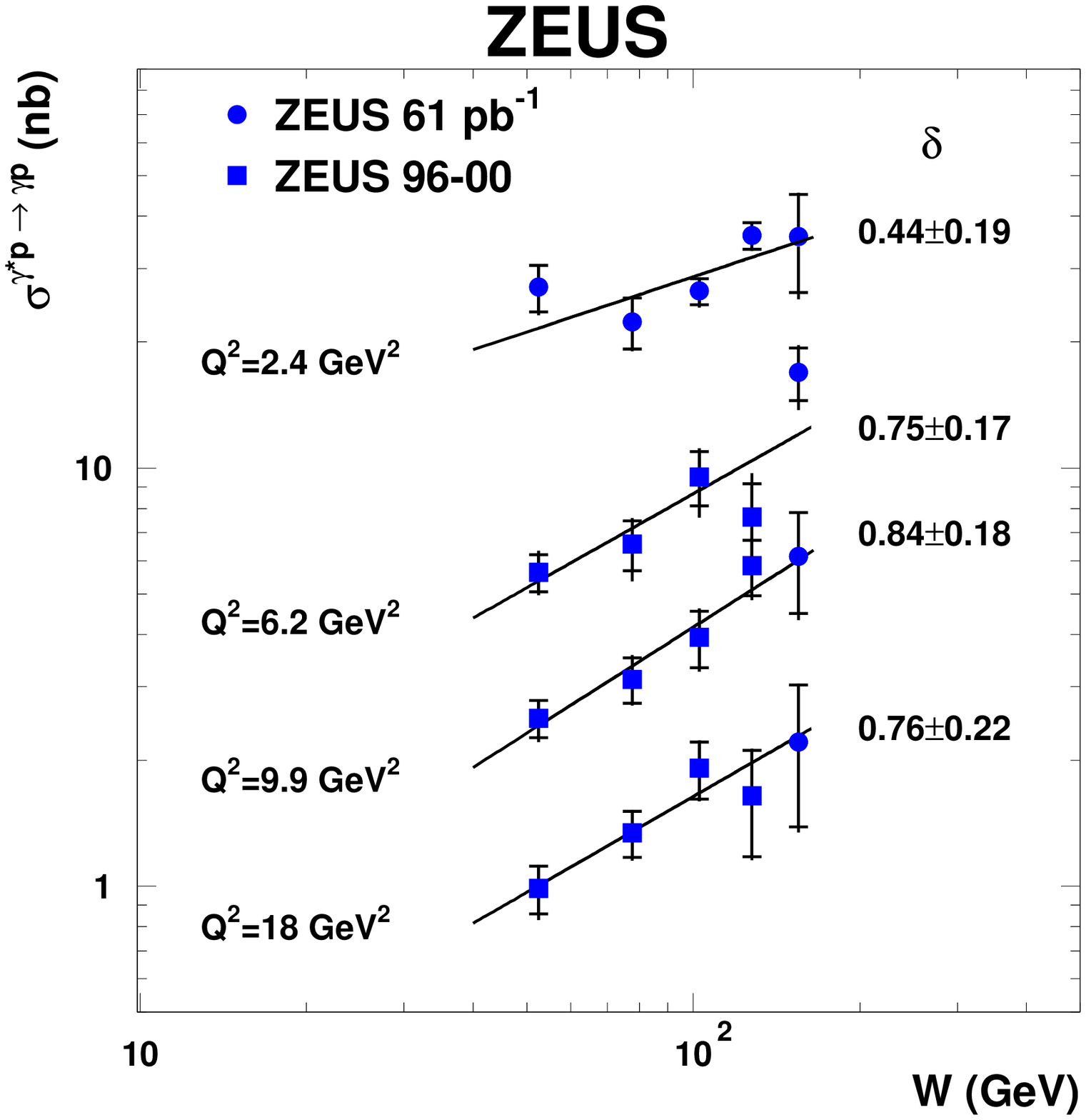}
\includegraphics[width=0.55\hsize]{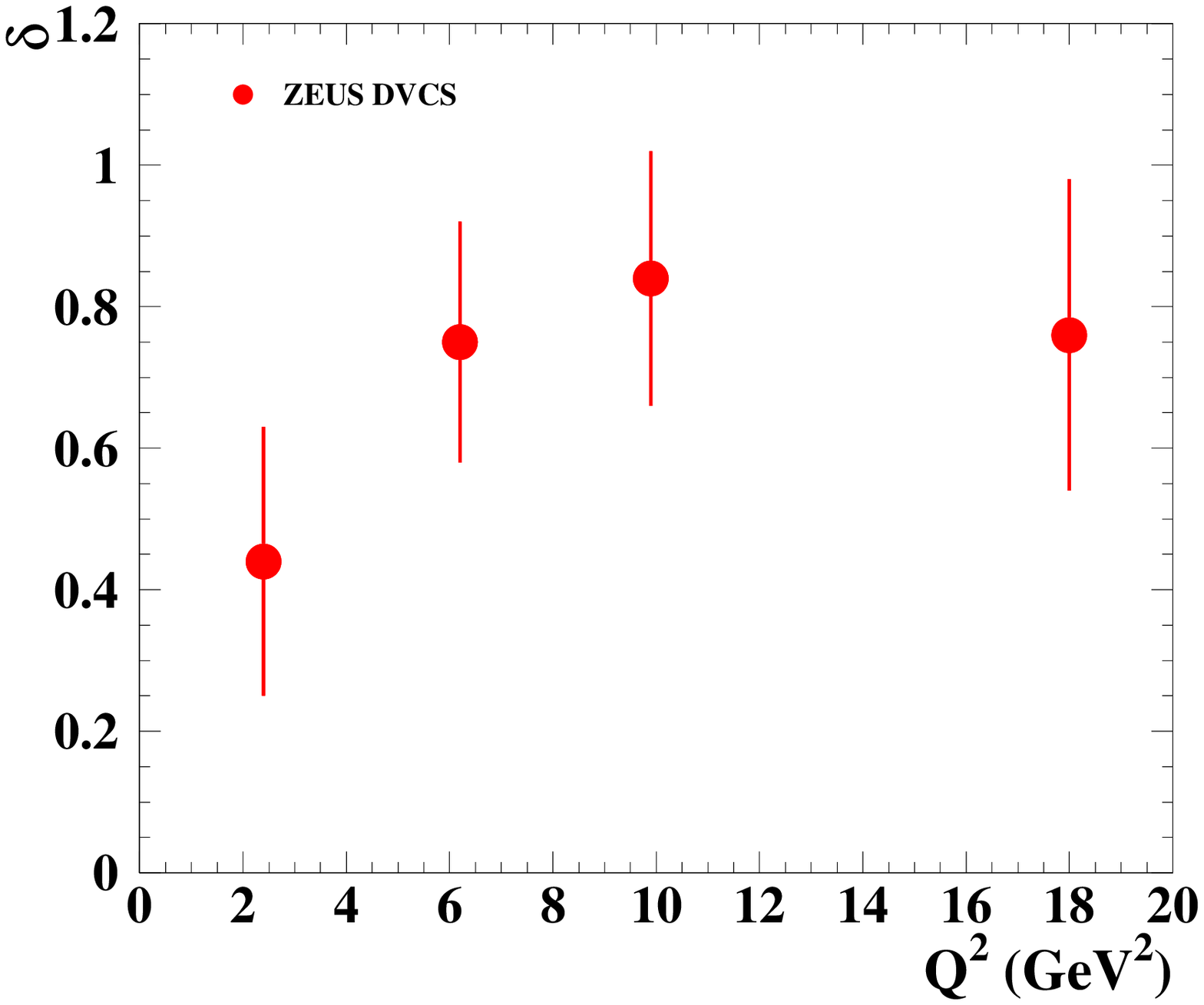}
\vspace*{-0.5cm}
\caption{
(Left-hand side) The $W$ dependence of the cross section for $\gamma^* p
\to \gamma p$ at different $Q^2$ values, measured by the ZEUS collaboration.
 The curves are fits of the form $\sim W^\delta$.(Right-hand side) The
 $Q^2$ dependence of $\delta$. }
\label{fig:dvcs-lps}
\end{figure}

The same situation is also true for $\sigma(\gamma^*p\to J/\psi p)$
and for $\sigma(\gamma^*p\to\gamma p)$, shown in
Fig.~\ref{fig:sig-dvcs}. 
For the latter, a recent measurement by the ZEUS collaboration, using
the leading proton spectrometer, is shown in
Fig.~\ref{fig:dvcs-lps}. Although the uncertainties on $\delta$ are
large also in this case, the trend of an increase with $Q^2$ is
apparent.

\begin{figure}[htb]
\begin{center}
\includegraphics[width=0.6\hsize]{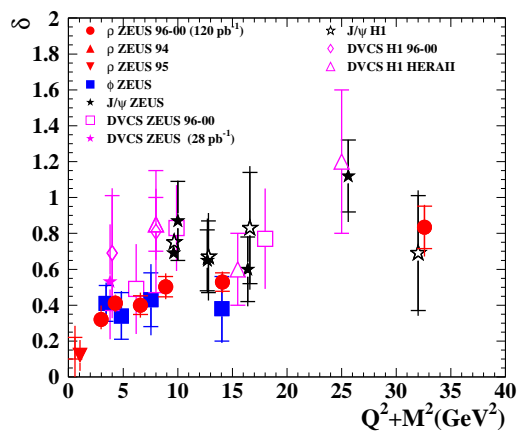}
\caption{
A compilation of all values of $\delta$ as a function of $Q^2+M^2$,
for exclusive electroproduction of $\rho^0, \phi, J/\psi$ and DVCS. }
\label{fig:delta09}
\end{center}
\end{figure}

A compilation of all values of $\delta$ for the different vector
mesons, $\rho^0, \phi, J/\psi$ and DVCS, is shown in
Fig.~\ref{fig:delta09}, as function of $Q^2+M^2$, where $M$ is the
mass of the vector meson. One sees an approximate universal behaviour,
showing an increase of $\delta$ as the scale becomes larger, in
agreement with the expectations mentioned above. The value of $\delta$
at low scale is the one expected from the soft Pomeron intercept,
while the one at large scale is in accordance with that expected from
the square of the gluon density. Note that the chosen scale of
$Q^2+M^2$ might not be the appropriate effective scale for all the
mesons.  For further discussion about the effective scale, see
section~\ref{sec:scale}.

\section{The ratio $\sigma(\gamma^*p\to Vp)/\sigma_{tot}(\gamma^*p)$}

As we saw in the previous section, the cross section for exclusive
electroproduction of vector mesons exhibits a rise with $W$, which
becomes steeper with increasing $Q^2$. It is of interest to compare the
$W$ dependence of this cross section to that of the total
$\gamma^*p$ one. To this end we study the ratio
\begin{equation}
r_V \equiv \frac{\sigma(\gamma^*p\to Vp)}{\sigma_{tot}(\gamma^*p)},
\end{equation}
for different vector mesons, as a function of $W$ for fixed
$Q^2$. Before presenting the data, we discuss the expectations for
$r_V$ using pQCD and Regge arguments.

\subsection{expectations for $r_V$ in pQCD}

In pQCD, the forward cross section for longitudinally polarized
photons is expected~\cite{brodsky} to behave as

\begin{equation}
\frac{d\sigma_L}{dt}|_{t=0} \propto
\frac{1}{Q^6}\alpha_S(Q^2)[xg(x,Q^2)]^2 ,
\end{equation}
where $x$ is the Bjorken scaling variable, $xg(x,Q^2)$ is the gluon
density in the proton and $t$ is the square of the
four-momentum-transfer at the proton vertex.

Assuming an exponential $t$ behaviour of the form $d\sigma_V/dt\sim
e^{bt}$, one gets the following expectation:

\begin{equation}
r_V \propto \left( 1+\frac{1}{R} \right) \frac{W^{2\lambda}}{b}
\approx \frac{W^{2\lambda}}{b} .
\label{eq:rvpqcd}
\end{equation}

In expression~(\ref{eq:rvpqcd}) we have used the fact that both the
gluon density distribution and the proton structure function have a
similar $x^{-\lambda(Q^2)}$ dependence and that the ratio $R$ of the
vector meson production cross section induced by longitudinal and
transverse virtual photons, $R=\sigma_L/\sigma_T$, increases with
$Q^2$ but is $W$ independent~\cite{zeusrho} and thus can be neglected.

\subsection{expectations for $r_V$ in the Regge approach}

Using Regge phenomenology arguments~\cite{regge}, one expects $\sigma_V
\sim W^{4(\apom(0)-1)}/b$ and $\sigma_{tot} \sim W^{2(\apom(0)-1)}$, where 
$\apom(0)$ is the intercept of the Pomeron trajectory, and therefore
\begin{equation}
r_V \propto \frac{W^{2(\apom(0)-1)}}{b},
\end{equation}
which is the same as~(\ref{eq:rvpqcd}), if we write $\lambda = \apom(0)-1$.

Both in pQCD and Regge approaches the ratio $r_V$ rises with $W$. The
$W$ dependence is not strongly affected by $b$ since both for the
exclusive electroproduction of $\rho^0$ and $J/\psi$ shrinkage was
found to be small~\cite{zeusrho,zjpsi}.

\subsection{Ratio - at what scale?}
\label{sec:scale}

\begin{figure}[htb]
\begin{center}
\includegraphics[width=0.6\hsize]{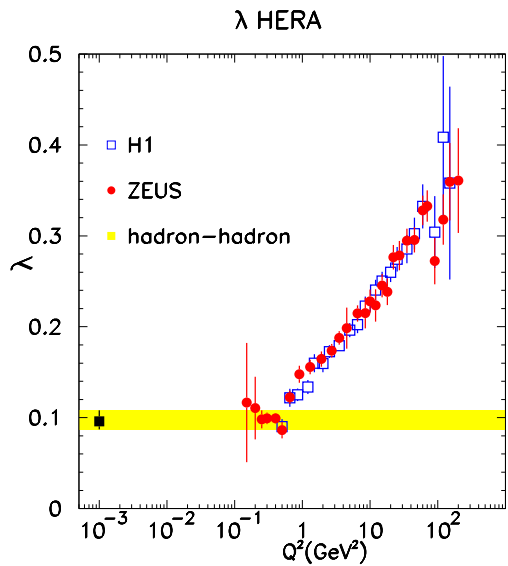}
\caption{
The values of $\lambda$, obtained by fitting the proton structure
function $F_2$ to the form $\sim x^{-\lambda}$, as a function of
$Q^2$. The shaded band shows $\apom(0)$-1 as obtained from
hadron-hadron total cross section data.
}
\label{fig:lam-hera}
\end{center}
\end{figure}

When calculating the ratio $r_V$ one has to ensure that both cross
sections are taken at the same hard scale, which we denote as
$Q^2_{eff}$. Clearly, for the total inclusive cross section, the
effective scale is $Q^2$. In Fig.~\ref{fig:lam-hera} one can see the
$Q^2$ dependence of $\lambda$ resulting from fitting the $F_2$ data in
the low-$x$ region ($x<0.01$) to the form $F_2 \sim x^{-\lambda}$.
For photoproduction and the low $Q^2$ region, the value of $\lambda$
is in good agreement with that expected from the Pomeron intercept
($\lambda=\apom(0)-1)$. Starting at about $Q^2>$1 GeV$^2$, the value
of $\lambda$ rises logarithmically with $Q^2$. It is of interest to
see if $r_V$ shows the expected $W^{2\lambda}$ behaviour.

\begin{figure}[htb]
\begin{center}
\includegraphics[width=0.4\hsize]{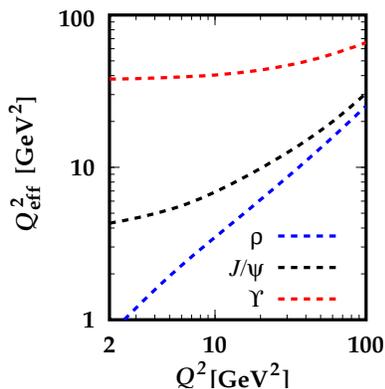}
\end{center}
\caption{The effective scale $Q^2_{eff}$ as a function of $Q^2$ for $\rho$, 
$J/\psi$ and $\Upsilon$.}
\label{fig:fks}
\end{figure}

It is not clear what is the effective scale
for exclusive vector meson electroproduction. One
suggested scale, originally for the $J/\psi$~\cite{ryskin}, is
\begin{equation}
Q^2_{eff} = \frac{Q^2 + M^2}{4}.
\label{eq:ryskin}
\end{equation}
where $M$ is the mass of the vector meson. Frankfurt, Koepf and
Strikman~\cite{fks} calculated the effective scale for $\rho, J/\psi$
and $\Upsilon$, shown in Fig.~\ref{fig:fks}, and find that their
effective scale for the $J/\psi$ and for the $\Upsilon$ are
significantly larger than that suggested by~(\ref{eq:ryskin}). 
One can parameterise~\cite{strikman} their effective scale for the
$\rho^0$ as
\begin{equation}
Q^2_{eff} = \left(\frac{Q^2}{2.65}\right)^{0.887}.
\label{eq:strikman}
\end{equation}

\begin{figure}[htb]
\begin{center}
\includegraphics[width=0.325\hsize]{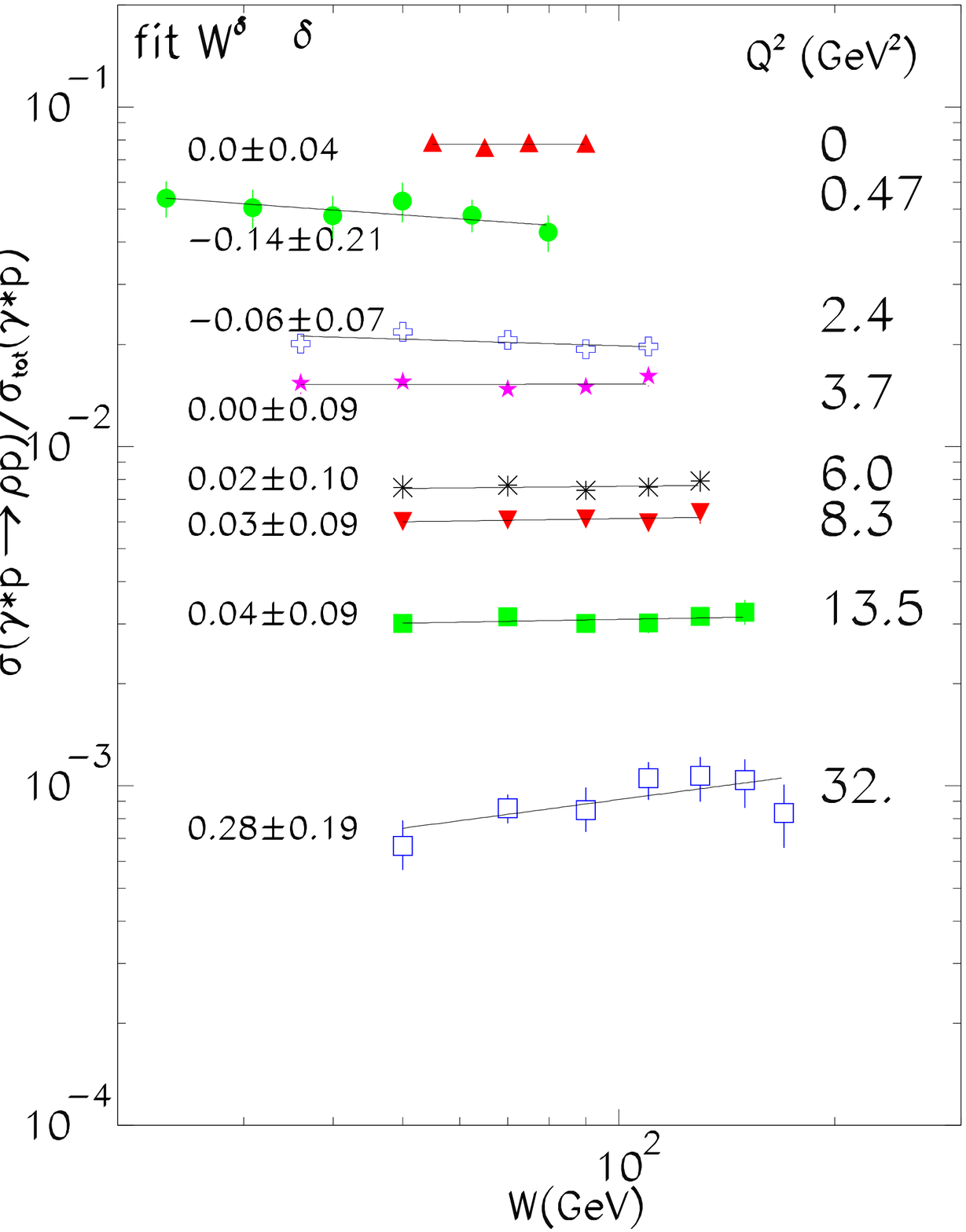}
\includegraphics[width=0.325\hsize]{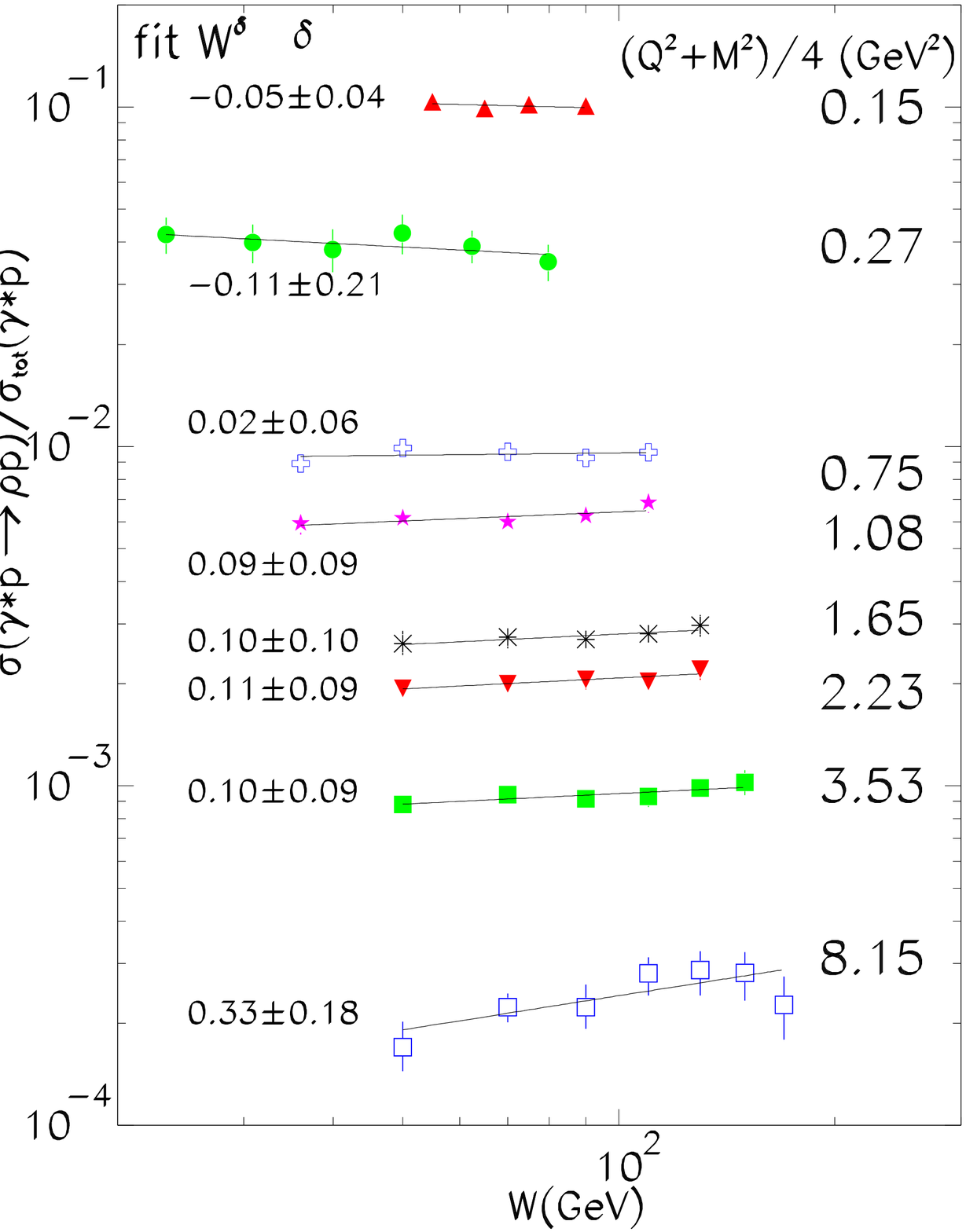}
\includegraphics[width=0.325\hsize]{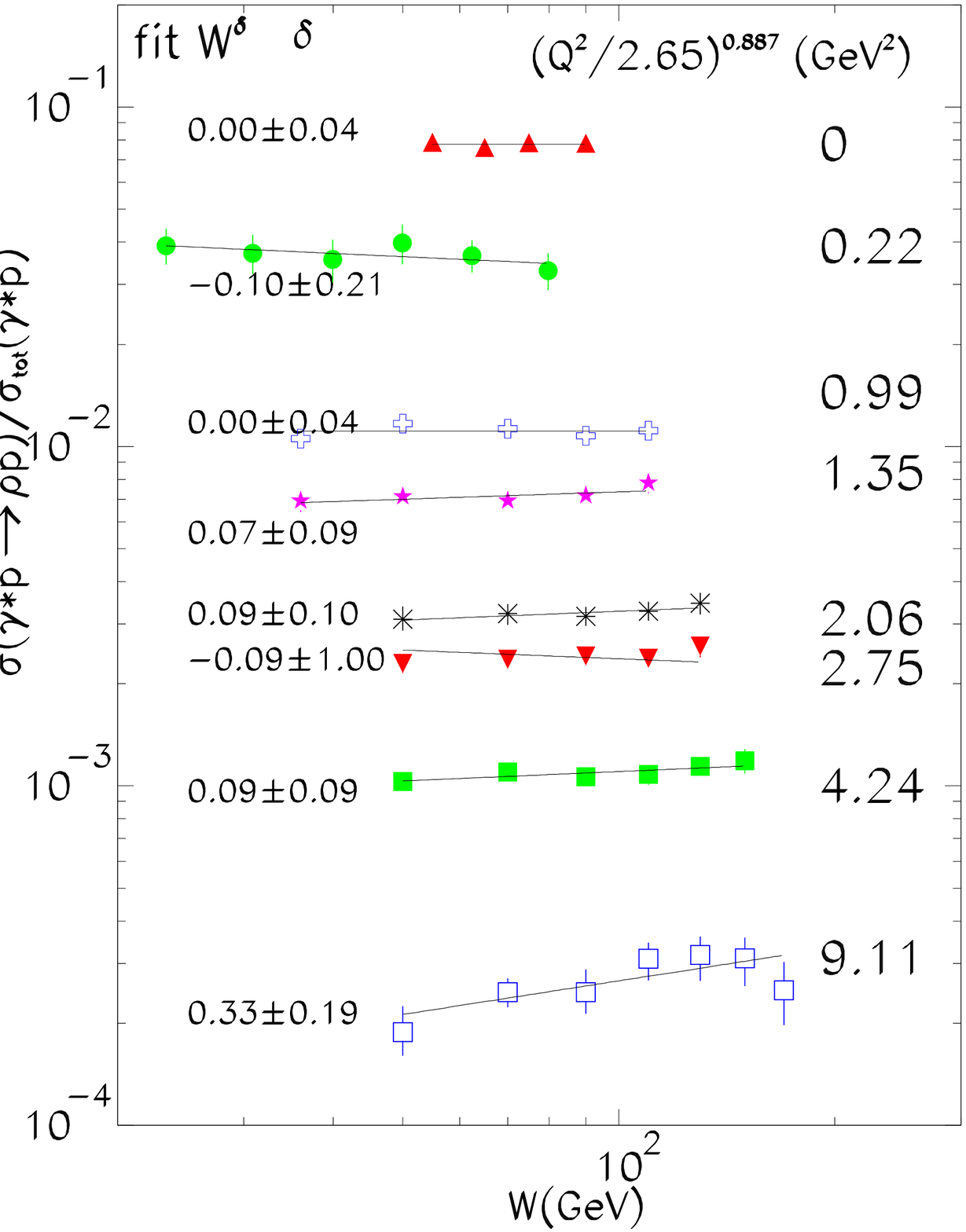}
\caption{
The ratio of $\sigma_{\rho^0}/\sigma_{tot}$ as a function of $W$ for
three choices of the effective scale; $Q^2_{eff}=Q^2$(left),
$Q^2_{eff}=(Q^2+M^2)/4$ (middle) and $Q^2_{eff}=(Q^2/2.65)^{0.887}$
(right). The lines are fit of the form $\sim W^\delta$ and the
resulting values of $\delta$ are given in the figures.  }
\label{fig:rho-tot1}
\end{center}
\end{figure}

As the most precise data at present are those of the exclusive
electroproduction of $\rho^0$~\cite{zeusrho}, we will use these data
to investigate the question of the effective scale.  The ratio
$\sigma(\gamma^* p \to
\rho^0 p)/\sigma_{tot}(\gamma^* p)$ is plotted in
Fig.~\ref{fig:rho-tot1} as a function of $W$, for different fixed
effective scales.
Three effective scales are used: $Q^2_{eff}=Q^2$,
$Q^2_{eff}=(Q^2+M^2)/4$, and $Q^2_{eff}=(Q^2/2.65)^{0.887}$. At each
fixed scale, the $W$ dependence is fitted to the form $\sim W^\delta$
and the value of $\delta$ is given in the figure. The ratio
$r_{\rho^0}$ seems to be constant with $W$ and shows a possible $W$
dependence only at the highest scale. We have shown earlier that this
ratio is expected to grow with $W$ like $W^{2\lambda}$ in both the
pQCD and the Regge approaches. This would thus indicate that the
$\lambda$ values obtained from this ratio are inconsistent with those
obtained in the inclusive total deep inelastic cross section case.

\begin{figure}[htb]
\begin{center}
\includegraphics[width=0.325\hsize]{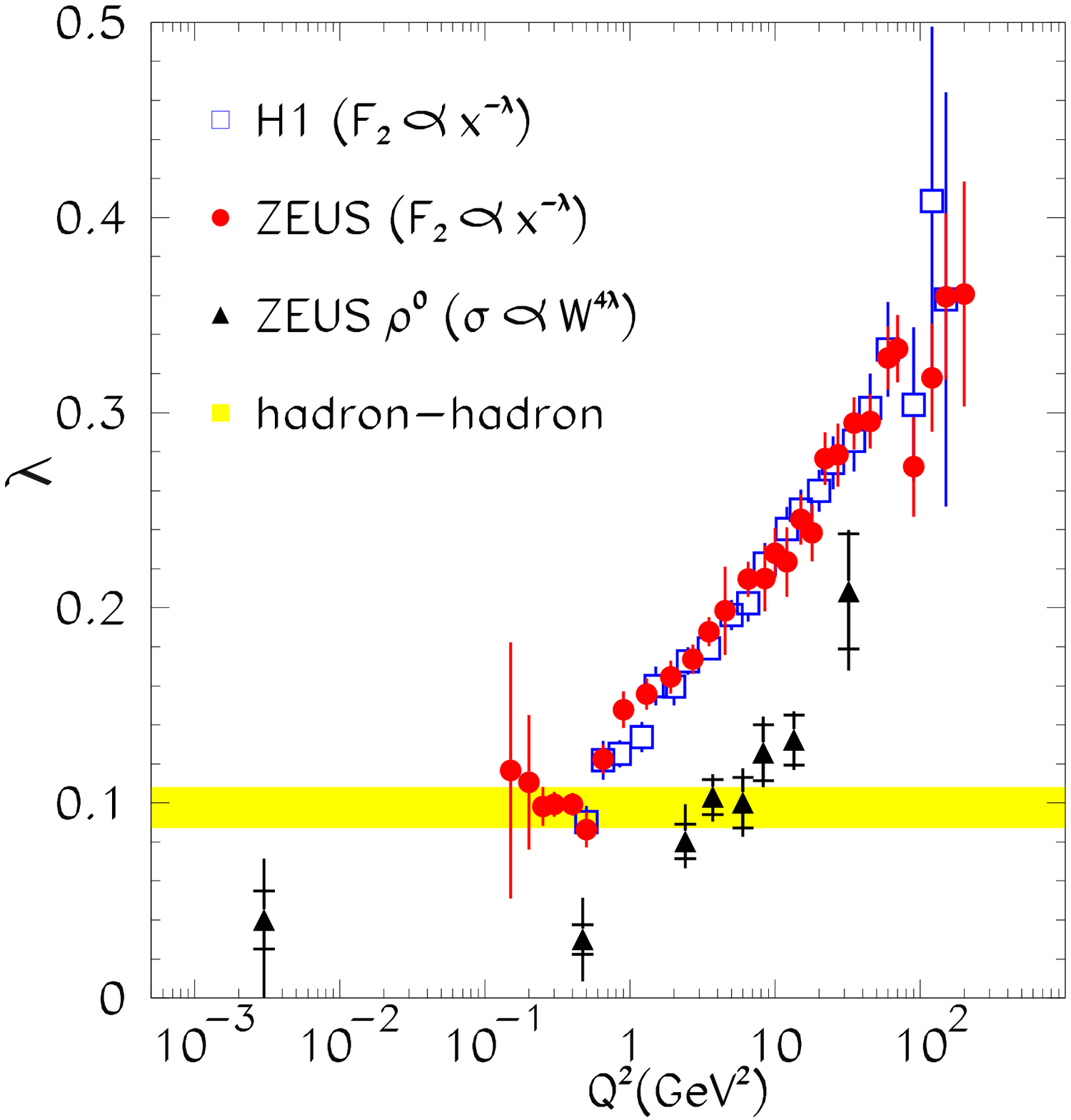}
\includegraphics[width=0.325\hsize]{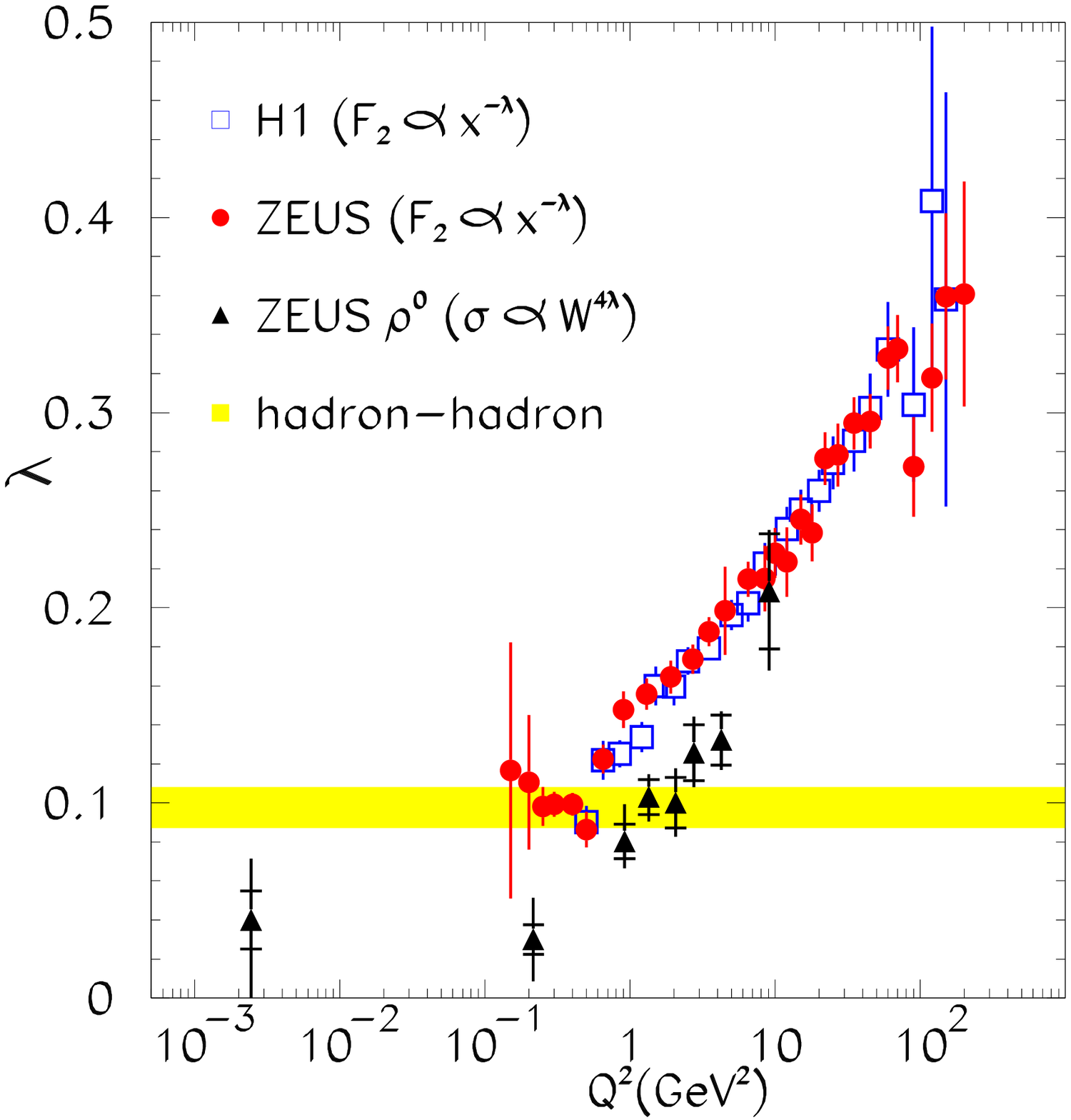}
\includegraphics[width=0.325\hsize]{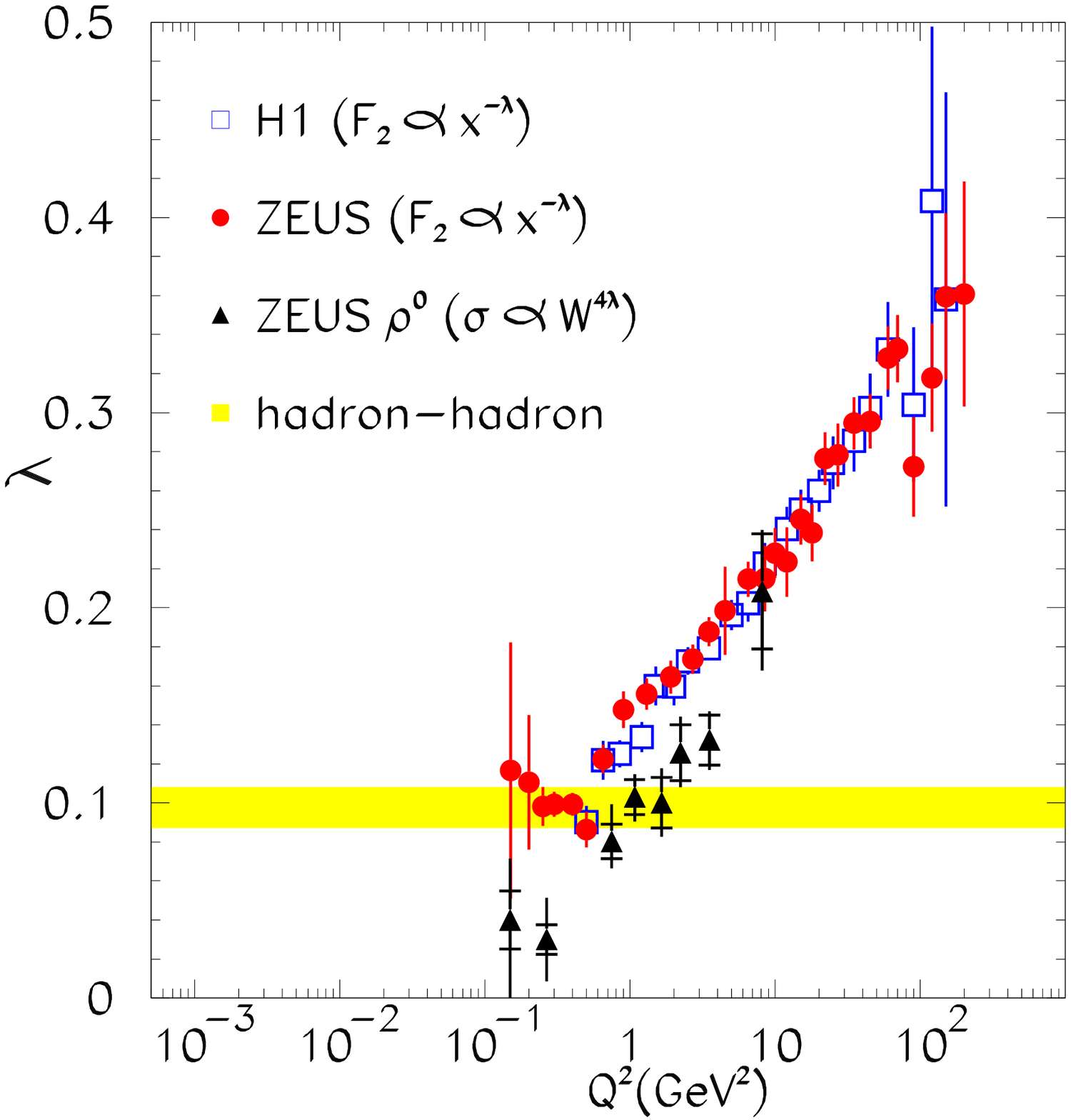}
\caption{
Comparison of the $\lambda$ values obtained from the exclusive
electroproduction of $\rho^0$ with those from the total inclusive
cross section, as a function of $Q^2$ for three choices of the
effective scale of the $\rho^0$; $Q^2_{eff}=Q^2$(left),
$Q^2_{eff}=(Q^2+M^2)/4$ (middle) and $Q^2_{eff}=(Q^2/2.65)^{0.887}$
(right).  }
\label{fig:lam-rho41}
\end{center}
\end{figure}

We can in fact compare directly the values of $\delta$ obtained from
the $W$ dependence of the $\rho^0$ cross section, shown on the
right-hand side of Fig.~\ref{fig:sig-rho}, to the $\lambda$ values
shown in Fig.~\ref{fig:lam-hera}, keeping in mind that
$\delta=4\lambda$. This is shown in Fig.~\ref{fig:lam-rho41}, for the
three effective scales mentioned above.
For all three scales, the $\lambda$ values obtained from the $W$
dependence of the $\rho^0$ are lower than those of the inclusive one
for all but the highest effective scale of the $\rho^0$.

\begin{figure}[htb]
\begin{center}
\hspace*{-0.5cm}
\includegraphics[width=0.5\hsize]{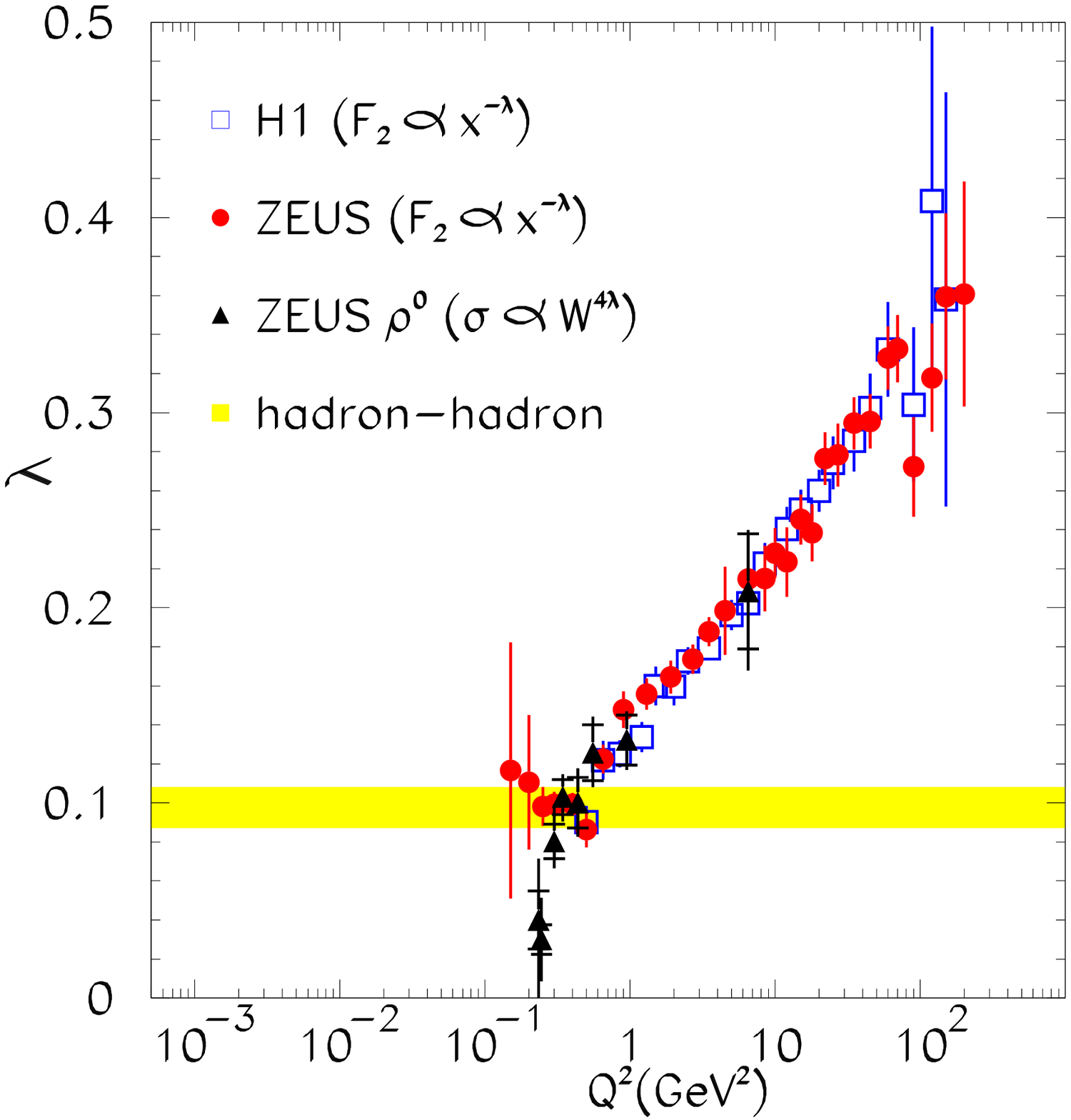}
\includegraphics[width=0.5\hsize]{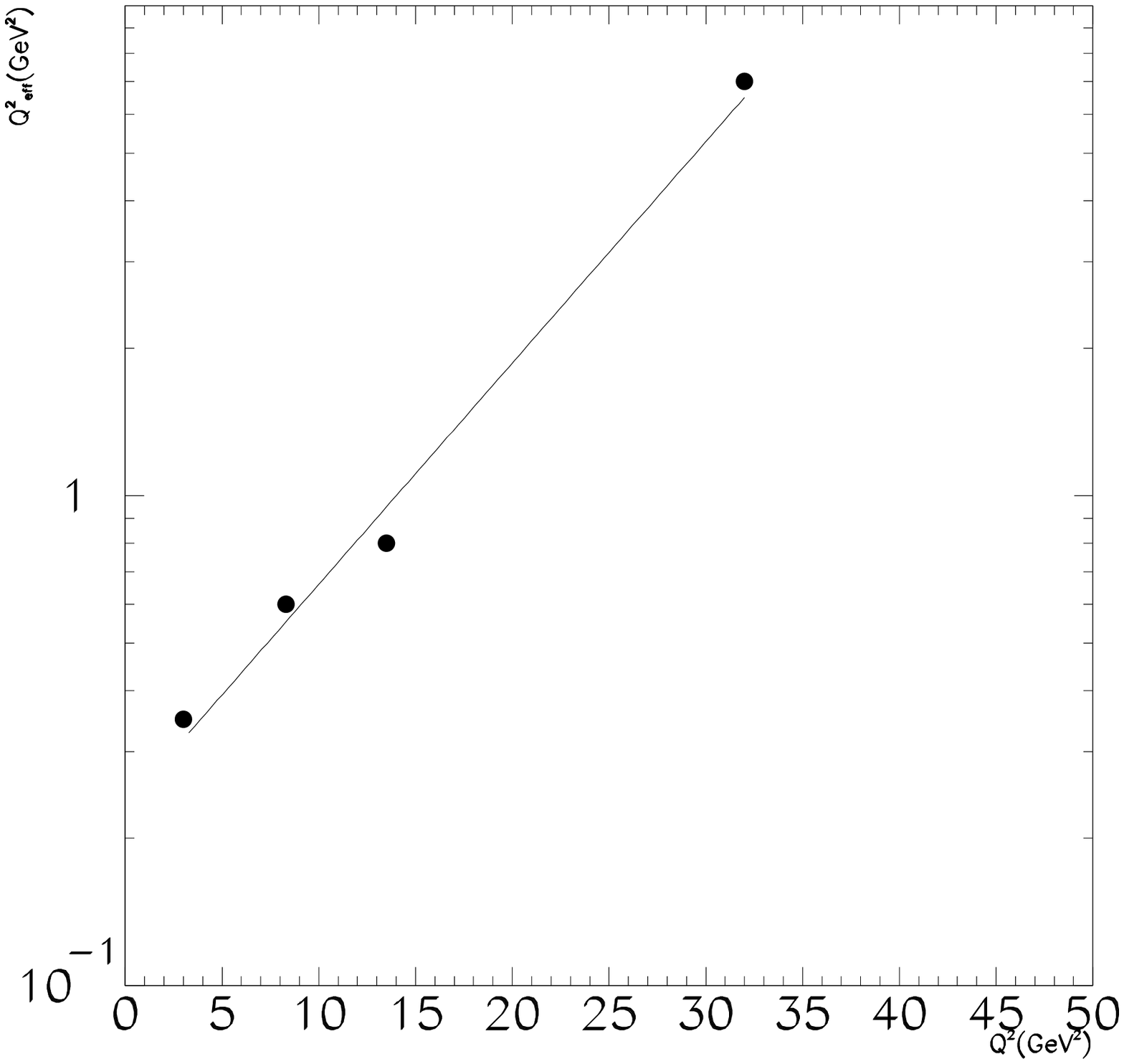}
\end{center}
\caption{(Left-hand side) The $\lambda$ values obtained from the exclusive
electroproduction of $\rho^0$ together with those from the total
inclusive cross section, as a function of $Q^2$. The effective scale
of $\rho^0$ was chosen so that the two $\lambda$ agree. (Right-hand
side) The effective scale of the $\rho^0$, $Q^2_{eff}$, as a function
of $Q^2$. The line is the result of an exponential fit to the data.
}
\label{fig:lam-rho42}
\end{figure}

One way to find experimentally the right effective scale is to force
the $\lambda$ from the $\rho^0$ to agree with that of the inclusive
data. This is shown on the left-hand side of Fig.~\ref{fig:lam-rho42}. 
The resulting values of $Q^2_{eff}$ are shown on the right-hand side
of Fig.~\ref{fig:lam-rho42}. The effective scale is much smaller than
($Q^2+M_{\rho})$/4 for low $Q^2$ values. An exponential fit to the
data, valid in the range where the measurements were done, results in
the following ad hoc parameterisation
\begin{equation}
Q^2_{eff} = 0.23 e^{0.1Q^2}.
\label{eq:al}
\end{equation}
The fact that the $Q^2_{eff}$ in the exclusive $\rho^0$
electroproduction is much smaller than $Q^2$ of the photon might be
due to the presence of the convolution of the soft $\rho^0$
wave-function and the small size longitudinal photon
wave-function. This is a clear sign of the interplay of soft and hard
physics~\cite{afs}.

One can revisit now the $r_{\rho^0}$ plot, using for the effective
scale of the $\rho^0$ as given in~(\ref{eq:al}). 
\begin{figure}[htb]
\begin{center}
\includegraphics[width=0.5\hsize]{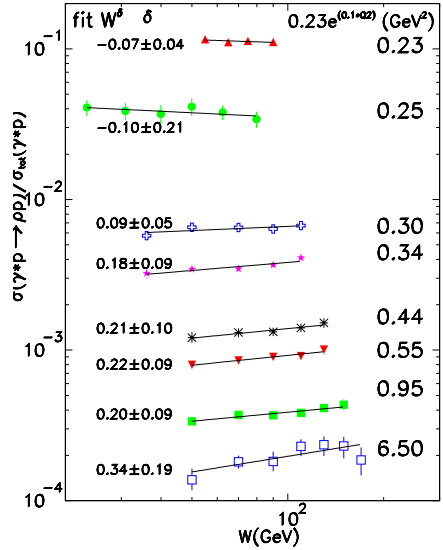}
\caption{
The ratio of $\sigma_{\rho^0}/\sigma_{tot}$ as a function of $W$ for
the effective scale $Q^2_{eff}=0.23 e^{0.1Q^2}$. The lines are fit of
the form $\sim W^\delta$ and the resulting values of $\delta$ are
given in the figure.  }
\label{fig:rho-tot2}
\end{center}
\end{figure} 
This is shown in Fig.~\ref{fig:rho-tot2}, where one can see now that
the ratio rises with $W$, as expected, even for the low effective
scales.

\begin{figure}[htb]
\begin{center}
\includegraphics[width=0.4\hsize]{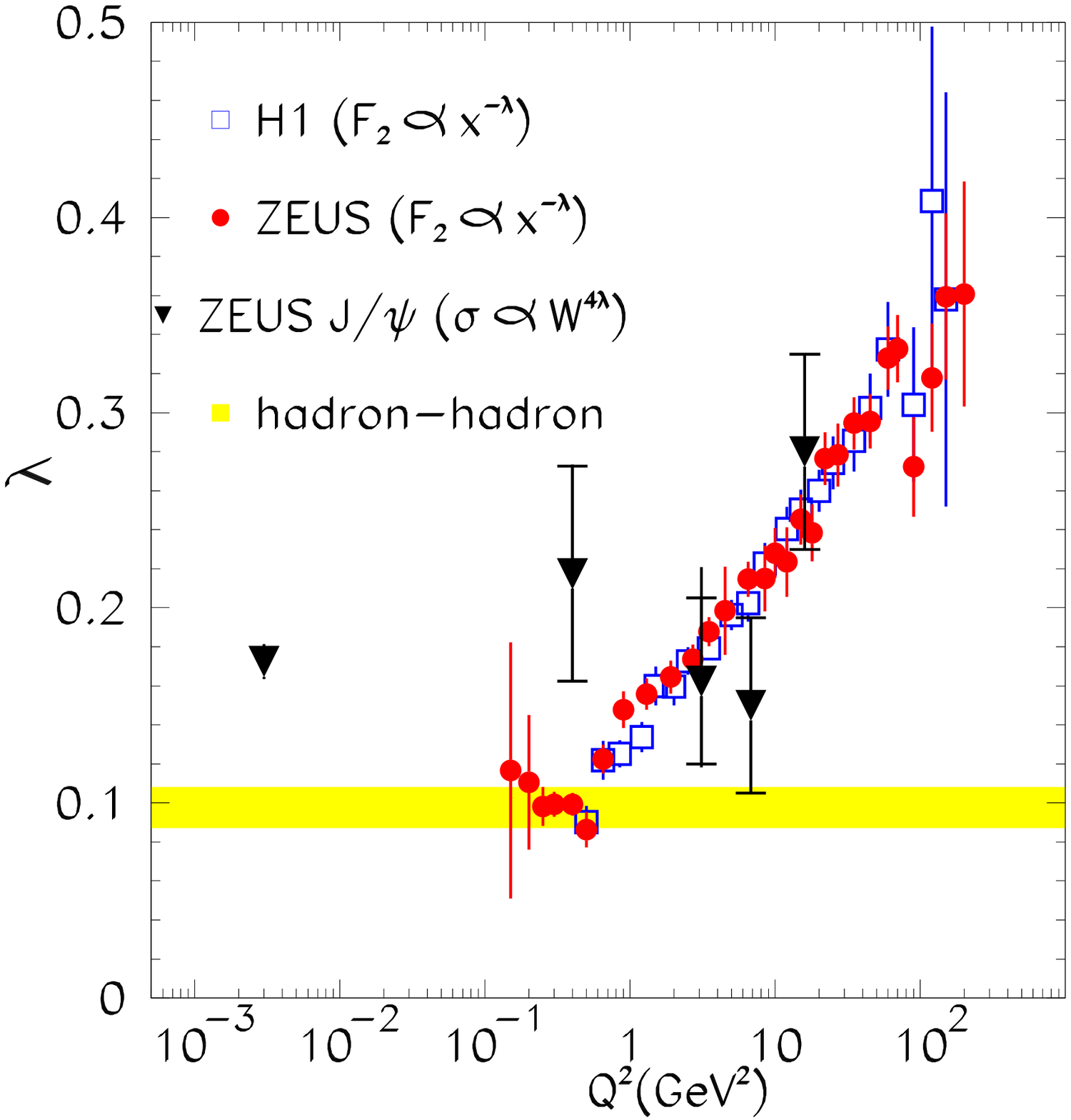}
\includegraphics[width=0.4\hsize]{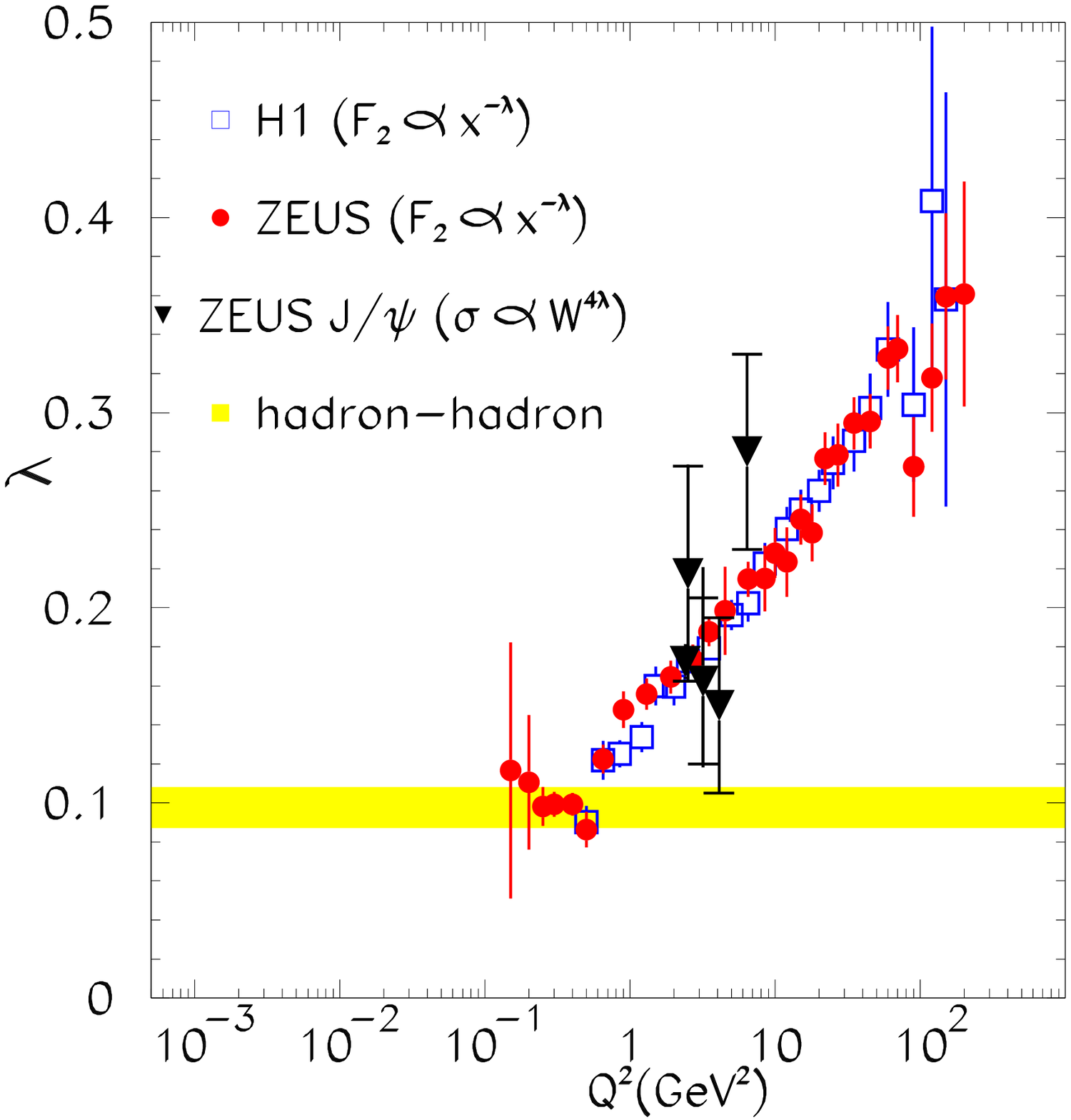}
\end{center}
\caption{
Comparison of the $\lambda$ values obtained from the exclusive
electroproduction of $J/\psi$ with those from the total inclusive
cross section, as a function of $Q^2$ for two choices of the
effective scale of the interaction; $Q^2_{eff}=Q^2$(left),
$Q^2_{eff}=(Q^2+M^2)/4$ (right).}
\label{fig:lam-psi}
\end{figure}

Unfortunately, the precision of the data of exclusive
electroproduction of $J/\psi$ does not allow a similar study. The
$\lambda$ comparison for the $J/\psi$ and inclusive data is shown in
Fig.~\ref{fig:lam-psi} for two effective scales, $Q^2$ and
$(Q^2+M^2)/4$. 
Clearly $Q^2_{eff}=Q^2$ is inappropriate in the low $Q^2$
region. However, the uncertainty of the data points of the exclusive
channel does not allow further conclusions.

In principle, the question of the effective scale should not be an
issue at all. If we were able to calculate QCD to all orders, we would
know exactly what is the right scale. However, as long as we do not
yet have a full calculation, precision measurements of exclusive
electroproduction of vector mesons would be helpful to resolve this
problem.

\section{$\lambda$ from inclusive diffraction}

\begin{figure}[htb]
\begin{center}
\includegraphics[width=0.5\hsize]{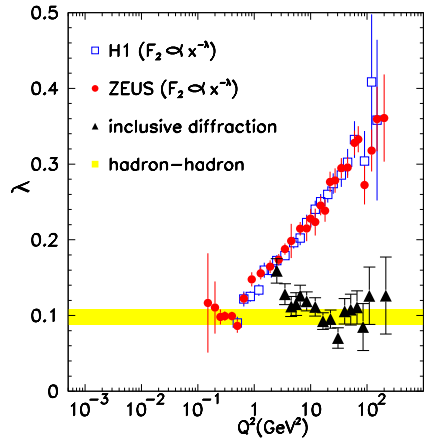}
\caption{
Comparison of the $\lambda$ values obtained from inclusive diffraction
with those from the total inclusive cross section, as a function of
$Q^2$.  }
\label{fig:lam-incl}
\end{center}
\end{figure}

Following the comparison of $\lambda$ from exclusive diffraction to
that from total inclusive reactions in the previous section, it is of
interest to make a similar comparison for inclusive diffraction. To
this end, the recent data~\cite{zeus-lrg} on $\apom(0)$ as a function
of $Q^2$ is used, with the relation $\lambda=\apom(0)-1$. The results
from the inclusive diffraction and the total inclusive reactions are
shown in Fig.~\ref{fig:lam-incl}.
The inclusive diffraction process shows a completely different
behaviour from that of the total inclusive one; the values of
$\lambda$ seem to be $Q^2$ independent and follow the expectations of
a soft process. This indicates that the source of the
large-rapidity-gap formation is a soft process.

One way to understand this feature is through the picture first
advertised by Gribov~\cite{gwee} and by Feynman~\cite{fwee} who use
the concept of 'wee' partons. The fact that one reaches high $W$
values gives enough time for the cloud of partons to develop from
'perturbative partons' to 'non-perturbative partons', the latter being
dressed large-size configurations, named as 'wee' partons. The large
size configurations lead to a soft process in the formation of large
rapidity gaps.

\section{Total photoproduction cross section at HERA}

\begin{figure}[htb]
\begin{center}
\includegraphics[width=0.5\hsize]{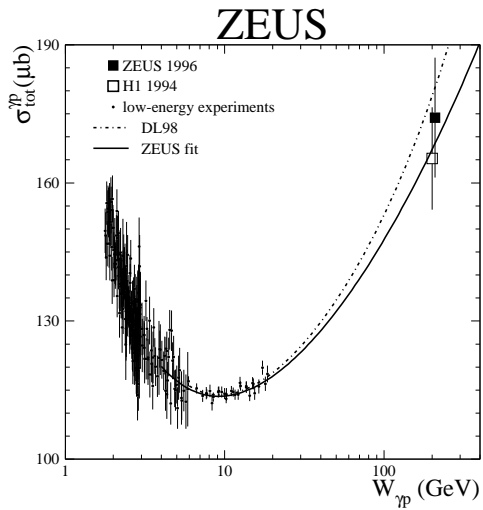}
\caption{
The photon-proton total cross section as a function of the
photon-proton center-of-mass energy.  }
\label{fig:sigtot-01}
\end{center}
\end{figure}

The first measurements~\cite{sigtot92,toth1-1} of $\sigma_{tot}(\gamma
p)$, though having large uncertainties, were enough to establish that
the photon behaves at large energies as a hadron. Later attempts to
improve the precision of the measurement~\cite{ztot,htot} showed that
while the statistical uncertainty can be much reduced, the systematic
uncertainty is still large (see Fig.~\ref{fig:sigtot-01}) and can not
allow a precise determination of the energy dependence.

During the last period of HERA running, the proton energy was changed
from 920 GeV (high energy run - HER) to 575 GeV (medium energy run -
MER) and to 460 GeV (low energy run - LER). The main purpose was to
measure the longitudinal structure function $F_L$ of the proton. In
addition, a special photoproduction trigger was implemented to extract
the value of $\lambda$ by measuring ratios of cross sections at
different $W$ values and thus minimizing the systematic uncertainties.

Assuming $\sigma \sim W^\delta$~\cite{ztot},
\begin{equation}
r=\frac{\sigma(W_1)}{\sigma(W_2)}=\left(\frac{W_1}{W_2}\right)^\delta \, .
\end{equation}
Experimentally,
\begin{equation}
\sigma = \frac{N}{A\cdot \cal{L}} \, ,
\end{equation}
where $A$, $\cal L$ and $N$ are the acceptance, luminosity and number
of measured events, respectively, and therefore
\begin{equation}
r=\frac{N_1}{N_2}\cdot \frac{A_2}{A_1} \cdot \frac{{\cal L}_{2}}{{\cal
L}_{1}} \, ,
\label{acc}
\end{equation}
where the index 1(2) denotes measurements performed at $W_1$
($W_2$). The acceptance for $\gamma p$ events at HERA depends mainly
on the detector infrastructure in the electron (rear) direction. If
the change in the $W$ value results from changing the proton energy,
the acceptance is likely to remain the same independent of $W$ and
the ratio of acceptances will drop out in Eq.~(\ref{acc}).

A preliminary measurement, using HER and LER data, was presented at
DIS08~\cite{dis08}. The value of $\delta$ as obtained from $r$ was
\begin{equation}
\delta = 0.140 \pm 0.014(\rm{stat.}) \pm 0.042(\rm{syst.}) 
\pm 0.100(\rm{6mT}),
\end{equation}
which translates into
\begin{equation}
\lambda = 0.070 \pm 0.007(\rm{stat.}) \pm 0.021(\rm{syst.}) 
\pm 0.050(\rm{6mT}).
\end{equation}
This result is consistent with earlier determinations of $\lambda$
(also denoted as $\epsilon$), however has the advantage of being
obtained from a single experiment and being model independent.

The statistical uncertainty will be improved in the future by
including the data taken with a third proton beam energy (MER). The
systematic uncertainty will improve by a better understanding and
calibration of the tagger of the scattered electron (6mT).

\section{Summary}

The HERA data are a good source to observe the interplay of soft and
hard dynamics, through the study of energy dependences of different
processes.

The process of exclusive electroproduction of vector mesons at high
scales is a good source to study perturbative QCD. It is important to
understand the issue of the effective scale. For the $\rho^0$, the
effective scale is much smaller than the $Q^2$ of the photon. Better
precision measurements are needed for the $\phi$, $J/\psi$ and DVCS to
get a determination of the effective scale in these processes.

There are plans to measure precisely the energy dependence of the
$\gamma p$ total cross section. This would set the baseline of soft
interactions to which $\gamma^* p$ results can be compared.

\section{Acknowledgments}
It is a pleasure to thank the organisers for an excellent and pleasant
meeting organised in memory of Prof. Jan Kwiecinski.

This work was supported in part by the Israel Science 
Foundation (ISF).

\end{document}